\documentclass[journal]{IEEEtran}
\ifCLASSINFOpdf
\else
\fi
\usepackage{amssymb}
\usepackage{amsmath}
\usepackage{algorithm}
\usepackage{algorithmic}
\usepackage{hyperref}
\usepackage{tikz}
\def\checkmark{\tikz\fill[scale=0.4](0,.35) -- (.25,0) -- (1,.7) -- (.25,.15) -- cycle;} 
\usepackage{graphicx}
\usepackage{booktabs}
\usepackage{xspace}

\usepackage{multirow}
\usepackage{makecell}

\usepackage{ mathrsfs }
\usepackage{graphicx, adjustbox}
\usepackage{array}
\usepackage{setspace}
\usepackage{enumitem}
\usepackage{breqn}
\usepackage{bm}
\usepackage{ dsfont }
\usepackage{ upgreek }
\usepackage{textcomp}
\usepackage{multirow}
\usepackage{algorithm}
\usepackage{algorithmic}
\usepackage{caption}
\usepackage{pgfplots}
\usepackage{tkz-euclide}
\usepackage{subcaption}
\usepackage{tikz}{\tiny }
\usepackage{balance}
\usepackage{cite}
\usepackage{ tipa }

\usepackage{fancyhdr}
\usepackage{ upgreek }
\usepackage{soul}
\usepackage{breqn}
\usepackage[english]{babel}
\usepackage{braket} 
\usepackage{bbm}

\addto\captionsenglish{}
\allowdisplaybreaks

\usepackage{amsthm}

\usepackage{fancyhdr}
\usepackage{ upgreek }
\usepackage{soul}
\makeatletter
\newcommand{\vast}{\bBigg@{4}}

\newcommand{\Vast}{\bBigg@{5}}
\begin{document}

\title{Towards Heterogeneous Quantum Federated Learning: Challenges and Solutions}

\author{Ratun Rahman, Dinh C. Nguyen, Christo Kurisummoottil Thomas, and Walid Saad, ~\IEEEmembership{Fellow,~IEEE} 

\thanks{Ratun Rahman and Dinh C Nguyen are with the Department of Electrical and Computer Engineering, University of Alabama in Huntsville, Huntsville, AL 35899, USA, emails: rr0110@uah.edu, dinh.nguyen@uah.edu} 
\thanks{Christo Kurisummoottil Thomas is with the Department of Electrical and Computer Engineering, Worcester Polytechnic Institute, Worcester, MA, 01609, and Walid Saad is with the Bradley Department of Electrical and Computer Engineering, Virginia Tech, Alexandria, VA, 22305, USA, emails: christokt@vt.edu, walids@vt.edu.}

}

\maketitle

\begin{abstract}
Quantum federated learning (QFL) combines quantum computing and federated learning to enable decentralized model training while maintaining data privacy.  QFL can improve computational efficiency and scalability by taking advantage of quantum properties such as superposition and entanglement.  However, existing QFL frameworks largely focus on homogeneity among quantum \textcolor{black}{clients, and they do not account} for real-world variances in quantum data distributions, encoding techniques, hardware noise levels, and computational capacity. These differences can create instability during training, slow convergence, and reduce overall model performance.
In this paper, we conduct an in-depth examination of heterogeneity in QFL, classifying it into two categories: data or system heterogeneity. Then we investigate the influence of heterogeneity on training convergence and model aggregation.  We critically evaluate existing mitigation solutions, highlight their limitations, and give a case study that demonstrates the viability of tackling quantum heterogeneity.  Finally, we discuss potential future research areas for constructing robust and scalable heterogeneous QFL frameworks.
\end{abstract}

\maketitle



 \section{Introduction}
 

\IEEEPARstart{Q}{uantum} machine learning (QML) \textcolor{black}{is a promising approach in machine learning} (ML) as it can process complex and large-scale data at an unprecedented speed by leveraging quantum phenomena such as superposition, entanglement, quantum parallelism, and quantum interference. QML integrates quantum physics with advanced computational techniques in ML across distributed quantum devices known as noisy intermediate-scale quantum (NISQ) devices~\cite{schuld2014quest}. However, in most conventional QML frameworks, the data is collected and processed on a central server, raising significant privacy concerns and exposing the data to various data-based attacks, even with quantum encoding~\cite{larasati2022quantum}. Furthermore, the large-scale nature of high-dimensional data transfer creates a high communications overhead, resulting in slower overall performance and scalability challenges~\cite{oh2020tutorial}. These limitations make QML unsuitable for practical deployment in scenarios that involve continuous data processing and sensitive data handling. 

\textcolor{black}{A promising approach to address the aforementioned challenge is through the use of quantum federated learning (QFL)~\cite{chehimi2022quantum}. QFL combines quantum computing with federated learning (FL) to perform ML tasks across distributed networks. In a QFL configuration}, multiple clients equipped with quantum devices conduct local data encoding and use quantum states and unique quantum features such as superposition and entanglement~\cite{chehimi2022quantum, ren2023towards}. Each client creates a local model by processing data in quantum form, which enables the use of quantum computational benefits such as faster processing rates and more effective handling of complex data sets. These local models are modified by sending the improved parameters back to a central server, aggregating the modifications.  The server uses complex quantum algorithms to conduct this aggregation, which are particularly intended to improve the collective learning process throughout the quantum network.  This type of model aggregation not only protects data privacy, but also uses quantum computing capabilities to improve learning outcomes, particularly when dealing with complex large-scale computational problems~\cite{ren2023towards,qiao2024transitioning}.

\begin{table*}[ht]
\footnotesize
    \centering
    \begin{tabular}{p{1.7cm}|c|p{1.8cm}|c|p{1cm}|p{.8cm}|p{8.5cm}}
    \hline
         \textbf{Reference} & \textbf{Year} & \textbf{Quantum components} & \textbf{QFL} & \textbf{Hetero-geneity} & \textbf{Case study} & \textbf{Contribution} \\
         \hline
         Larasati et al.,~\cite{larasati2022quantum} & 2022 & - & \checkmark & - & - & Provided a basic overview and the potential of QFL.\\
         \hline
         Ren et al.,~\cite{ren2023towards} & 2023 & - & \checkmark & - & - & Detailed a thorough taxonomy for QFL, noted its challenges and future direction.\\
         \hline
         Qiao et al.,~\cite{qiao2024transitioning} & 2024 & \checkmark & \checkmark & - & - & Provided a complete step-by-step procedure to QFL, identified challenges and future direction. \\
         \hline
         Chehimi et al.,~\cite{chehimi2023foundations} & 2024 & \checkmark & \checkmark & \checkmark & - & Examined the key components of QFL and briefly introduced the challenges of heterogeneity on QFL. However, it did not explain the overall challenges and solutions of non-independent and identically distributed (non-IID) QFL environments.\\
         \hline
         \textcolor{black}{This paper} & 2025 & \checkmark & \checkmark & \checkmark & \checkmark & \textcolor{black}{Provide in-depth exposition of the challenges of heterogeneity in QFL, potential solutions as well as the limitations of the existing method for heterogeneous QFL networks. A case study is developed to show how to overcome the heterogeneity created by quantum noise.}\\
         \hline
    \end{tabular}
    \caption{Comparison of existing works on QFL and new contributions of this article.}
    \label{tab:intro}
\end{table*}

However, despite such promising results, \textcolor{black}{ existing QFL frameworks~\cite{larasati2022quantum,ren2023towards,qiao2024transitioning,chehimi2023foundations} } mainly assume homogeneous clients, ignoring the inherent heterogeneity that exists in real-world quantum systems.  In practice, QFL clients have substantially distinct quantum data distributions, encoding strategies~\cite{larose2020robust}, hardware noise levels, and computational abilities, resulting in differences in local model training and global aggregation.  This variability causes model divergence, slow convergence, and suboptimal learning performance, making generalization difficult across different quantum devices.  Variations in the depth of quantum circuits, decoherence rates, and quantum gate fidelities also exacerbate learning differences between clients.  As a result, tackling heterogeneous QFL is critical for achieving robust, scalable, and practical federated learning~\cite{jose2022error}. 

\textcolor{black}{The main contribution of this paper is to fill the gap by providing an in-depth exposition of the challenge of heterogeneity in QFL over quantum networks.} Specifically, we investigate the nature of heterogeneity in QFL, how it affects training convergence and model performance, \textcolor{black}{and mitigation strategies to tackle the effects of heterogeneity.} We divide heterogeneity into two types: data heterogeneity and system heterogeneity, and we investigate the distinct issues that each presents in quantum networks.  Furthermore, we critically examine existing mitigation strategies, emphasizing their limitations and tradeoffs. Finally, we provide a case study that shows the advantages and feasibility of using methods that tackle quantum heterogeneity. The contributions of this article compared to the state-of-the-art are summarized in Table~\ref{tab:intro}.






\section {Fundamentals of QFL}
\subsection{Background}
\textcolor{black}{
\textbf{Quantum bits.} A quantum bit (qubit) is the fundamental unit of quantum computation.  A qubit, unlike classical bits, can hold both 0 and 1 at the same time, a feature known as superposition.  When two or more qubits interact, they can become entangled, which means their states are connected even when separated, allowing powerful computations~\cite{qiao2024transitioning}. This enables quantum systems to process a large number of possibilities simultaneously, providing considerable speedups over traditional approaches.}

\textcolor{black}{\textbf{Quantum gates.} Quantum gates are the basic operations that alter the states of qubits. Unlike conventional logic gates, they are reversible and can handle multiple possibilities at once. Pauli-X (a quantum variant of NOT gate), Hadamard (used for creating superposition), and CNOT (used for entanglement) are some of the most common gates used in QFL and quantum algorithms.} 

\textcolor{black}{
\textbf{Quantum gates.} A quantum layer is a series of quantum operations performed on qubits that is commonly used in variational quantum circuits (VQCs).  It is made up of parameterized gates (e.g., RX, RY, RZ) that rotate qubits and entangling gates (e.g., CNOT) that connect qubits to use superposition and entanglement.  These layers function as quantum analogs of neural network layers, enabling efficient encoding and learning of complex data patterns with fewer parameters.
}

\textcolor{black}{
\textbf{Quantum measurements.} Quantum measurement is how we obtain results from a quantum system.  A qubit exists in a variety of states, but measurement converts it to a determinate value, either 0 or 1, which we can use for classical methods such as gradient updates~\cite{qiao2024transitioning}.
}

\subsection{QFL Working Procedure}
\subsubsection{Quantum Encoding}
\textcolor{black}{Quantum encoding is the process of transforming classical data into quantum states using quantum gates for quantum computation}~\cite{chehimi2023foundations}.  Common encoding methods include \textit{basis encoding}, which directly maps classical bits to  $|0\rangle$ and  $|1\rangle$ states; \textit{amplitude encoding}, which represents data in quantum state amplitudes for compact storage; \textit{phase encoding}, which embeds information in the phase of qubits; and \textit{entanglement encoding}, which captures complex correlations using entangled states~\cite{larose2020robust}.  Different applications can benefit from each technique, and appropriate encoding guarantees that quantum circuits can handle classical data efficiently~\cite{qiao2024transitioning}. 

\subsubsection{Local Model Training}
In QFL, quantum local model training involves executing a quantum or hybrid quantum-classical model on each client separately before aggregating updates globally~\cite{oh2020tutorial}. It can be done in several distinct ways:  VQCs, which use parameterized quantum gates optimized via classical gradient descent; QNNs~\cite{schuld2014quest}, in which quantum layers replace classical ones to learn complex patterns; hybrid models that employ quantum encoding for feature extraction and traditional deep learning for optimization~\cite{larose2020robust}; and quantum kernel learning, which maps data to high-dimensional quantum spaces to improve classical models such as support vector machines~\cite{oh2020tutorial}. Each approach compromises between computational efficiency and noise resistance, ensuring effective training in distributed quantum clients~\cite{jose2022error}.

\subsubsection{Quantum Model Sharing}
Quantum model sharing in QFL involves transmitting trained quantum models between clients and a central server for global aggregation, which can be done using classical or quantum channels~\cite{chehimi2022quantum}. \textit{The classical channel} approach extracts and converts the features of the quantum model, such as the rotational angles of the quantum gates in VQCs, into classical data prior to transmission.  This approach is practical and compatible with current networks, but removes quantum correlations such as entanglement.  In contrast, \textit{quantum channels} allow for direct transfer of quantum states through quantum teleportation or quantum secure communication, which preserves quantum coherence. \textcolor{black}{However, they require a consistent entanglement distribution and high-fidelity quantum connections, making it more resource-intensive}.  

\subsubsection{Quantum Model Aggregation}
Quantum model aggregation in QFL integrates locally trained quantum models into a global model while accounting for quantum state coherence, entanglement, and noise shifts between clients~\cite{chehimi2022quantum}. It uses classical parameter aggregation, in which trainable quantum parameters (for example, gate angles in VQC) are extracted, translated to classical data, and averaged using methods such as federated averaging (FedAvg)~\cite{gurung2024personalized}. However, this method lacks quantum correlations.  Alternatively, quantum state aggregation maintains quantum features by directly integrating trained quantum states with entanglement-based methods or quantum interpolation. This results in improved coherence, but requires high-fidelity quantum communication and error correction.  Depending on hardware capabilities, quantum state aggregation provides better performance in completely quantum networks, whereas conventional aggregation is more practical for near-term quantum systems~\cite{chehimi2023foundations}. 

\vspace{-3mm}\subsection{Heterogeneity in QFL}
\textcolor{black}{QFL presents distinctive types of heterogeneity that extend beyond conventional FL, due to the inherent variability of quantum devices and hardware.  Devices vary in qubit count, coherence time, error rates, and gate integrity, directly impacting local training performance~\cite{shi2024personalized, gurung2024personalized}.  Unlike classical FL, where heterogeneity arises primarily from data distribution or computation capacity, QFL encounters inconsistencies in the encoding of quantum data and noise levels among clients~\cite{larose2020robust}.  These quantum-specific restrictions are not taken into consideration by classical FL approaches, which are intended to handle non-IID data or computational imbalance. As a result, they are inadequate for stabilizing and synchronizing training in QFL systems~\cite{qiao2024transitioning}.}
We can separate this heterogeneity into two \textcolor{black}{groups}: data heterogeneity and system heterogeneity. 

\section{Data Heterogeneity in QFL}
\textcolor{black}{In QFL, data heterogeneity refers to differences in the representations of quantum data between clients. It implies that various QFL clients may represent or process the same classical data differently due to differences in quantum encoding or hardware, making it challenging to successfully align and aggregate models throughout the network.}

{\color{black}
\textbf{How Data Heterogeneity in QFL Differs from Classical FL?}
In classical FL, clients have non-IID sample distributions and uneven compute power; however, all updates share the same Euclidean parameter space~\cite{qiao2024transitioning}. In contrast, data heterogeneity in QFL stems from the physics of the Hilbert space itself.} First, local encodings can convert an identical classical vector to non-orthogonal quantum states. However, because amplitudes and phases differ between clients, a naïve parameter average is theoretically meaningless when updates have incompatible bases. Second, the no-cloning theorem limits the ability to share an obtained quantum state, which means clients must transmit either low-fidelity transported states or classical shadows, introducing representation-dependent heterogeneous noise with no classical equivalent. Third, each parameterized quantum circuit (PQC) can entangle qubits according to its own topology; therefore, aggregating models that act on various tensor-product factors is significantly more difficult than combining conventional networks, which never experience tensor misalignment. Fourth, similar statistical skews appear as state-specific fidelity loss because decoherence links closely to the selected encoding (amplitude, phase, or basis)~\cite{larose2020robust}, making noise-agnostic solutions like FedProx infeasible. Finally, each diagnostic measurement compresses the quantum state; thus, even calculating local statistics disturbs the distribution being studied, while conventional histogramming is non-destructive.

\subsection{Heterogeneous Quantum Encoding}
Heterogeneous quantum encoding addresses situations in which federated clients use distinct quantum encoding methods to represent local data. \textcolor{black}{This scenario results in} differences in the way quantum information is structured and processed.  
Clients can utilize basis, amplitude, phase, or entanglement-based encoding, resulting in different representations throughout the system. Heterogeneous quantum data can emerge even when all federated clients \textcolor{black}{employ} the same encoding method, such as amplitude encoding, due to differences in data pretreatment, normalization, and quantum circuit design.  Even with a consistent encoding approach, discrepancies in classical data distributions before encoding may manifest in distinct quantum state distributions between clients~\cite{larose2020robust,gurung2024personalized}.

\textbf{Effects.} Heterogeneous quantum encoding leads to inconsistencies in quantum state representations, which have a negative impact on training stability and model convergence in QFL.  Even with the same encoding strategy, such as amplitude encoding, differences in input data scaling, feature distributions, and local quantum hardware characteristics can result in distinct constructed quantum states between clients~\cite{qiao2024transitioning}.  This results in divergent feature spaces, where local models extract various quantum representations, making it more challenging to create a globally consistent model.

\subsection{Multimodal Data Across Quantum Devices} 
In QFL, multimodal data is generated when clients manage varied inputs such as quantum states, entangled pairs, and measurement outcomes, alongside classical forms, including text, audio, and images~\cite{qiao2024transitioning}.  Each modality adds complimentary information, allowing QFL to use quantum processing capability for high-dimensional data while still including traditional modalities for deeper learning. Clients' sensing and preprocessing capabilities, as well as modality type, vary, resulting in heterogeneity.  One client may produce high-resolution quantum-enhanced visuals and precise state measurements, but another may just give text or coarse quantum outputs~\cite{qu2023qnmf}.  Such discrepancies restrict data fusion and model integration because modality-specific noise and format incompatibilities spread throughout global training.


\textbf{Effects.} The diversity of multimodal data between QFL clients complicates data integration and analysis, posing significant challenges for model training and overall performance.  Clients with varying types and qualities of data modalities contribute unevenly to the learning process, potentially skewing the global model toward data-rich or high-quality modes.  Such disparities impair the model's ability to generalize effectively across various quantum and classical datasets, reducing accuracy and operational efficiency in practical applications. In addition, differences in data integration methods and computational demands when processing such hybrid datasets often result in increased communication overhead and computational costs~\cite{qu2023qnmf}.  These factors have a negative effect on the synchronization and scalability of the QFL system.  

\section{System Heterogeneity in QFL}
\textcolor{black}{System heterogeneity in QFL refers to variances in quantum hardware between clients.  It implies that not all quantum clients can perform computations equally; some may have fewer qubits or greater noise levels, resulting in longer training, less accurate updates, or even the inability to do certain quantum operations consistently.}

\subsection{ Heterogeneous PQC Across Local QML Models}
A PQC is a quantum model comprised of gates controlled by adjustable parameters, commonly implemented as rotations around the $x$, $y$, or $z$ axes~\cite{qiao2024transitioning}.  By altering these parameters, PQCs may convert classical data to quantum states, explore high-dimensional Hilbert spaces, and extract patterns from measurement results.
In QFL, architectural heterogeneity occurs when clients use PQCs with varying depths or structural complexity~\cite{gurung2024personalized}.  These inequalities are mostly due to hardware variability, which includes variances in qubit availability, coherence durations, and gate quality.  Clients with limited resources may use shallow PQCs with fewer layers, but sophisticated devices can support deeper, more expressive circuits~\cite{rahman2025towards}.  This mismatch hampers global aggregation since parameters learnt from circuits with different capacities may not map directly across clients.

\textbf{Effects.} Heterogeneous PQC architecture across QFL clients has a major impact on the overall training process, resulting in significant differences in computing capability, model flexibility, and convergence.  Clients with fewer parameterized layers encounter limitations in describing sophisticated quantum data and complex correlations, resulting in less expressive local models.  Furthermore, changes in PQC complexity affect the global aggregation stage, since discrepancies in parameter dimensions prevent direct parameter averaging and increase communication costs.  Such disparities may require the use of specialized aggregating methods or approximation procedures, which may reduce overall training efficiency. Furthermore, structural heterogeneity presents fairness and equality problems, since clients with simpler PQCs may continuously provide less significant updates, biasing the global model toward participants with additional quantum resources~\cite{gurung2024personalized,shi2024personalized}.

\subsection{ Varying Number of Qubits across quantum devices}
QFL processes data in quantum bits (qubits), which are fundamentally different from classical bits.  While a classical bit is bound to two states (0 or 1), a qubit $|\psi\rangle$ can exist in a superposition, written as \(|\psi\rangle = \alpha |0\rangle + \beta |1\rangle, \quad 
    |\alpha|^2 + |\beta|^2 = 1,\) enabling more information encoding and parallelism~\cite{qiao2024transitioning}.
The number of accessible qubits differs across clients due to hardware technology, resource capacity, and noise resilience~\cite{jose2022error}.  Current devices have limited qubits, which are susceptible to decoherence due to environmental interference and need tight operating conditions~\cite{rahman2025towards}.  This unpredictability causes imbalances in local training capacity and global aggregate in QFL systems.

\textbf{Effects.} The difference in qubits in various QFL clients can considerably influence the FL training procedure. For example, clients equipped with fewer qubits have lower computational power, resulting in the implementation restriction of quantum circuits of higher complexity. In addition, qubits are used for quantum encoding~\cite{larose2020robust}, in which classical data is encoded into quantum data. Therefore, clients with fewer qubits can not adequately represent high-complexity data, which may lead to poor performance. It can also result in communication overhead, as clients with different qubits will cause inconsistent parameter size and quantum state dimension between devices~\cite{rahman2025towards}.

\subsection{Inherent Quantum Noise}
Inherent quantum noise occurs as a result of quantum decoherence, gate noise, and measurement irregularities in quantum systems, which vary between quantum devices~\cite{qiao2024transitioning}. As QFL includes several clients using distinct quantum processors, each of which experiences unique noise patterns, their local model updates become inconsistent~\cite{jose2022error}. 

\textbf{Decoherence.} Decoherence happens when qubits lose their quantum state as a result of environmental interactions, preventing superposition or entanglement.  Its rate is dependent on hardware specifications, cooling efficiency, and noise shielding, resulting in varied levels of stability among devices.  In QFL, such discrepancies cause clients with higher decoherence to provide noisier and less reliable parameter updates, diminishing global accuracy and slowing convergence.  Because quantum circuits rely on coherence for accurate computing, uneven decoherence causes unpredictability in local training and demands regular error correction.  This mismatch hampers federated aggregation because varied decoherence levels prevent synchronization and consistent global model performance.

\textbf{Gate noise.} Quantum gates conduct calculations on qubits, but gate noise is introduced by hardware faults, control imperfections, and external interference.  Devices vary in gate fidelities, causing certain clients to complete circuits more accurately than others.  In QFL, this leads to unequal local training, with cleaner updates from high-fidelity devices and error-prone contributions from noisier ones.  Gate noise, in addition to compute-capability differences in classical FL, is a significant source of heterogeneity that affects global performance and entanglement-based operations~\cite{jose2022error}.

\textbf{Quantum measurements.} Quantum gates use qubit manipulation to conduct calculations, but gate noise is introduced by hardware defects, control imperfections, and external interference.  Devices vary in gate fidelities, causing certain clients to complete circuits more accurately than others.  In QFL, this leads to unequal local training, with cleaner updates from high-fidelity devices and error-prone contributions from noisier ones.  Gate noise, in addition to compute-capability differences in classical FL, is a significant source of heterogeneity that affects global performance and entanglement-based operations~\cite{jose2022error}.

{\color{black}
\section{Mitigation Strategies for Heterogeneous QFL}
To address heterogeneity in QFL, we categorize mitigation options into four categories: encoding-level, model-architecture, hardware-aware, and noise-resilient solutions. 
\subsection{Encoding-Level Mitigations}
Data heterogeneity in QFL is more complex than in traditional FL.  While clients in QFL can transfer the same classical input to non-orthogonal quantum states based on encoding selection, normalization, or device noise, clients in FL share a Euclidean parameter space even when the data is not IID. 
 Direct heterogeneity of quantum states is further hindered by the no-cloning theorem, and naïve averaging is rendered incorrect by this Hilbert space mismatch.

\textbf{Encoding Harmonization.} To align distributions, classical inputs can be standardized before encoding, and synthetic states can be created using simulators or GANs.  However, augmentation must avoid non-physical states.

\textbf{Encoding-Aware Weighting.} When harmonization is not feasible, clients are weighted based on the similarity of their state $\rho_i$ to a reference $\rho_g$ as \(
    w_i = \frac{\exp(-\alpha \, d(\rho_i, \rho_g))}
    {\sum_{j=1}^N \exp(-\alpha \, d(\rho_j, \rho_g))},\)
where $d(\cdot,\cdot)$ is the quantum distance metric.  Clients with encodings similar to $\rho_g$ have a bigger influence on aggregation~\cite{gurung2024personalized}. 

To summarize, QFL's encoding-level mitigations tackle issues such as encoding-dependent noise and Hilbert space incompatibility that are not present in conventional FL.  Maintaining consistency among diverse customers involves weighting, harmonization, and continuous modification~\cite{rehman2023dawa}.

\subsection{Model-Architecture Strategies}
System heterogeneity in QFL is caused by clients with different PQC depths or qubit counts. QFL aggregation can be challenging when PQCs have mismatched structures or clients have restricted qubits, unlike conventional FL, where models have the same parameter dimensions.

\textbf{Layer-Wise PQC Aggregation.} To align diverse PQCs, only parameters from layers shared by all clients are aggregated as \(
    \theta_g^l = \frac{1}{|C_l|} \sum_{i \in C_l} \theta_i^l,\)
where $C_l$ represents the collection of customers with depth $\geq l$.  This prevents dimension mismatches during global updates~\cite{rehman2023dawa}.


\textbf{Qubit-Aware Embedding.} Clients with fewer qubits may contain their states in a larger Hilbert space prior to aggregation as \(
    \tilde{\rho}_i = U_i \rho_i U_i^\dagger, \quad 
    U_i : \mathbb{C}^{2^{q_i}} \to \mathbb{C}^{2^{q_g}},\)
where $q_i$ is the client’s qubit count and $q_g$ the global maximum.

\textbf{Circuit Compression.} Low-resource devices can participate without a complex circuit thanks to lightweight approximations (gate pruning, variational approximation)~\cite{rehman2023dawa}.

In summary, model-architecture mitigation techniques address structural incompatibilities particular to quantum systems.  Techniques like layer-wise aggregation, qubit-aware embeddings, and circuit compression allow diverse clients to communicate while maintaining quantum expressivity.

\subsection{Hardware-Aware Mitigations}
QFL clients often have varying hardware quality, coherence times, and available resources.  Quantum devices, unlike conventional FL, may have insufficient qubits or unstable gate fidelities, which can lead to biased global models.

\textbf{Hybrid Quantum–Classical Integration.} Low-resource clients can delegate large and complex computing to classical layers while maintaining quantum encoding and feature extraction.  This hybrid architecture allows for broad involvement without requiring deep PQCs on all devices~\cite{oh2020tutorial}.

\textbf{Personalized Synchronization.} Clients just synchronize globally compatible parameters, such as entangling layers, instead of requiring homogeneous circuits. The local update rule for regularization is calculated as \(
    \boldsymbol{\omega}_{i}^{t+1} = \boldsymbol{\omega}_{i}^t - \eta \Big( g_i^t + \lambda(\boldsymbol{\omega}_i^t - \boldsymbol{\omega}_g^t) \Big)\),
where the learning rate is $\eta$, the gradient is $g_i^t$, the global parameters are $\boldsymbol{\omega}_g^t$, and the regularization factor is $\lambda$. This factor balances customization and alignment~\cite{gurung2024personalized}.

\textbf{Fairness-Aware Weighting.} To prevent high-capacity devices from dominating, clients can scale their contributions based on effective capacity, such as qubit count $q_i$ or gate fidelity $\phi_i$ as \(
    w_i = \frac{q_i \cdot \phi_i}{\sum_{j=1}^N q_j \cdot \phi_j}.\)
This guarantees that updates accurately reflect both data quality and hardware dependability.

In summary, hardware-aware mitigations address device-level quantum restrictions, including qubit limits and gate integrity, unlike conventional FL.  Hybrid integration, selective synchronization, and fairness-aware weighing enable heterogeneous devices to interact while maintaining overall performance.

\subsection{Noise-Resilient Strategies}
Quantum devices vary in decoherence rates, gate errors, and measurement accuracy.  In QFL, hardware noise directly affects quantum state development, leading to inconsistent or biased local updates. This differs from conventional FL in which noise is often associated with stochastic optimization.

\textbf{Noise-Aware Aggregation.} Client updates can be weighted based on inverse noise variance, with steady devices contributing more as \(
    \theta_g = \frac{\sum_{i=1}^N \frac{1}{\sigma_i^2}\theta_i}
    {\sum_{i=1}^N \frac{1}{\sigma_i^2}},\)
where $\sigma_i^2$ is computed based on the fidelity or mistake rate of client $i$~\cite{larasati2022quantum}.

\textbf{Sporadic Participation.} To prevent unstable updates, only clients that fulfill the validation criterion $\tau$ can join the aggregation as \(
    i \in \mathcal{C}^{(t)} \iff \mathcal{A}_i^{(t)} \geq \tau,\)
where $\mathcal{A}_i^{(t)}$ represents the local accuracy in round $t$.  This adaptive participation minimizes error propagation during noise spikes~\cite{rahman2025sporadic}.


In summary, noise-resilient QFL techniques address decoherence, gate infidelity, and measurement errors, in addition to classical variance reduction.  Weighting, sporadic participation, and noise-tolerant VQAs work together to give resilience against diverse quantum noise.
}

\section{A Case Study: \textcolor{black}{Tackling Data Heterogeneity and Quantum Noise Using SPQFL}}
We present a case study that \emph{operates} the mitigation principles discussed in this work by proposing a \textit{sporadic personalized quantum federated learning (SPQFL)} protocol~\cite{rahman2025sporadic}.  In real‐world deployments, existing QFL frameworks suffer from (i) \emph{quantum noise}---stemming from heterogeneous device quality and decoherence susceptibility---and (ii) \emph{non‑IID data distributions} across clients~\cite{qiao2024transitioning}.  SPQFL jointly tackles these challenges and, as shown later, delivers consistent accuracy gains over contemporary base

\subsection{System Model}
{\color{black}
\textbf{Problem formulation.} We examine a QFL framework that consists of a single quantum server and a collection of distributed quantum devices. Each device trains a local quantum neural network (QNN) model using stochastic gradient descent (SGD) over multiple global rounds.} {\color{black} The global objective is to minimize the average empirical risk across all participating clients as

\begin{equation}
    \min_{\boldsymbol{\omega}} \frac{1}{N} \sum_{i=1}^N \mathbb{E}_{(x,y)\sim \mathcal{D}_i} \mathcal{L}(f(x;\boldsymbol{\omega}_i), y),
\end{equation}
where $N$ is the number of clients, $\mathcal{D}_i$ is the local dataset for client $i$, $\boldsymbol{\omega}_i$ are the local parameters, and $\mathcal{L}$ is the cross-entropy loss function.}

{\color{black}
\textbf{Datasets and preprocessing.} We examine SPQFL using four benchmark datasets: MNIST, Fashion-MNIST, CIFAR-100, and Caltech-101.  All photos are normalized with zero mean and unit variance.  Amplitude encoding converts classical inputs to quantum states.  Data is divided into non-IID client datasets, with each client receiving samples from a subset of classes to represent actual heterogeneity.  In each dataset, 80\% of samples are used for training and 20\% for testing.}

{\color{black}
\textbf{Local training.} Each device performs many local training epochs during a global cycle.  A QNN model is built with PQCs that include rotation gates ($R_x, R_y, R_z$) and entangling CNOT gates.  Training is done on random mini-batches of size $32$.  The Adam optimizer starts with a learning rate of $0.001$ and decays by $0.9$ per 10 global rounds.
}
{\color{black}
Local model parameters are updated as:
\begin{equation} \label{eqnlabelnew1}
    \omega_{n,k}^{t+1} = \omega_{n,k}^t - \eta \Big( g_{n,k}^t 
    + \lambda(\omega_{n,k}^t - \omega_k^t) \Big),
\end{equation}
where $\eta$ is the learning rate, $g_{n,k}^t$ is the local gradient, 
$\lambda$ is the personalization regularization coefficient, and 
$\omega_k^t$ denotes the global model parameters at round $t$.
}

{\color{black}
\textbf{Global aggregation.} After local training, each client communicates the modified parameters to the server.  To prevent noisy or unstable updates, only clients with validation accuracy above $\tau$ are included in the global aggregate (sporadic learning).  The global model is updated by taking the weighted average of selected client parameters and redistributing them for the following cycle as
\begin{equation} \label{eq: fedavg_personalized}
    \boldsymbol{w}_{k+1} = \frac{1}{N} \sum_{n=1}^{N} \boldsymbol{w}_{n,k}^{T}.
\end{equation}

\begin{figure}
    \centering
    \includegraphics[width=0.99\linewidth]{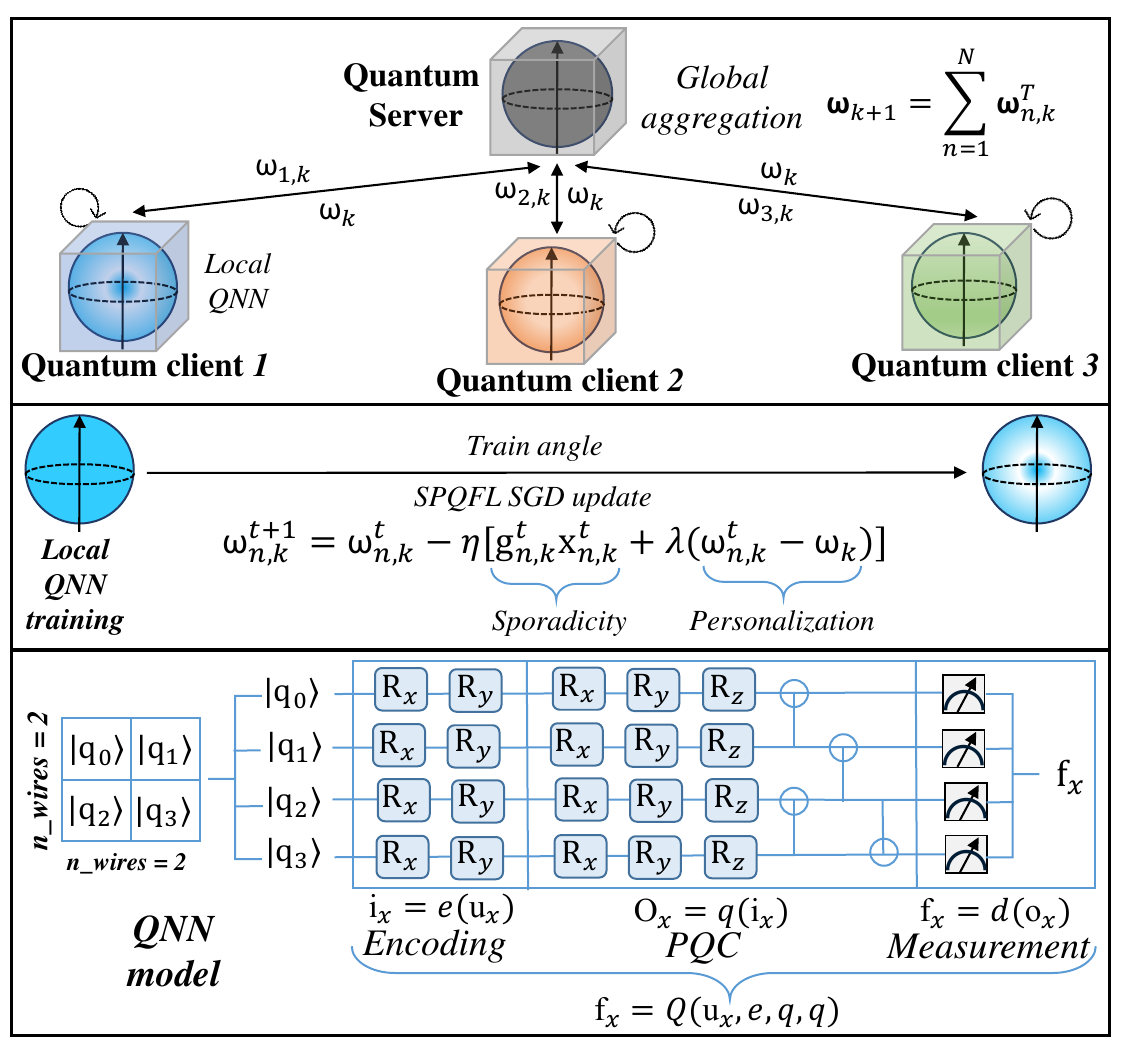}
    \caption{\textcolor{black}{Proposed \textit{SPQFL} architecture in which a set of distributed quantum devices collaborate with a quantum server to train a shared QML model. }}
    \label{fig: overview}
\end{figure}

\textbf{Simulation environment.} Quantum experiments are conducted using simulators due to the limited scalability of current technology. 
 We use the PennyLane and Qiskit Aer simulators for circuit execution.  We use the Lindblad master equation to describe real-world hardware errors, including noise channels.
 \begin{itemize} 
 \item \textit{Amplitude damping noise} with damping rate $\gamma$,
 \item \textit{Phase damping noise} with dephasing probability $p$, 
 \item \textit{Thermal relaxation noise} with relaxation times $T_1 = 50 \,\mu s$ and $T_2 = 70 \,\mu s$. 
 \end{itemize}
 These values correspond to features of near-term superconducting quantum devices.  The evaluations are conducted on a high-performance computer cluster equipped with NVIDIA RTX 4090 GPUs and 64 GB of RAM in the Linux operating system.}


\textcolor{black}{Our proposed SPQFL overview is described in Fig.~\ref{fig: overview}, where a group of distributed quantum clients collaborate to train a shared QML model under the coordination of a central quantum server.  Each client has a local QNN whose parameters are updated using the SPQFL variation of stochastic gradient descent, which includes two essential terms: a sporadic component and a personalization component.  The sporadicity phrase refers to device-specific noise and irregular communication, while the personalization term allows each client to remain adaptable to its unique data distribution. 
Each client transmits its model parameters $\omega_{n,k}$ to the quantum server for global aggregation after local training on encoded quantum data \(\omega_{k+1} = \sum_{n=1}^{N} \omega_{n,k}^{T}\), as seen in the top panel and forms the global model. The local QNN training dynamics are shown in the center panel, which also shows how each client uses both random and customized gradients to update parameters.  Lastly, the lower panel describes the internal structure of the QNN model itself. It uses a \textit{PQC} $o_x = q(i_x)$ to process classical inputs into quantum states $i_x = e(u_x)$, then \textit{measured} to produce classical outputs $f_x = d(o_x)$.  $f_x = Q(u_x, e, q, d)$ is an end-to-end flow that illustrates how SPQFL combines measurement, quantum processing, and encoding within the federated optimization loop.}

{\color{black} We summarize our proposed \textit{SPQFL} methodology explained above in Algorithm \ref{alg:sporadic_fl}. The server starts by initializing a global model and goes through \(K \) communication rounds (Line \ref{line:rounds}).  In each round, a subset of customers is chosen and given the current global model.  Each client performs local training in parallel (Line \ref{line:client_loop}).  For each \(T \) local step (Line \ref{line:local_loop}), the client first computes its local gradient, which incorporates noise from quantum computing (Line \ref{line:gradient}). To limit the influence of noise, the client computes a noise-aware scaling factor using an exponential decay function (Line \ref{line:weight}).  This component influences the gradient step, resulting in a noise-controlled model update with model personalization (Line \ref{line:update}).  Following local training, each client sends its final model to the server, which combines the results to update the global model.  This approach enables the system to dynamically suppress unreliable updates, increasing resilience to noise.

\begin{algorithm}[t]
\color{black}
\caption{Sporadic Quantum Federated Learning}
\label{alg:sporadic_fl}
\begin{algorithmic}[1]
\STATE \textbf{Input:} Learning rate \( \eta \), number of rounds \( K \), local steps \( T \), noise sensitivity \( \gamma \), number of clients \( N \)
\STATE \textbf{Initialize:} Global model \( \omega_0 \)
\FOR{each round \( k = 1 \) to \( K \)} \label{line:rounds}
    \STATE Server sends \( \omega_k \) to selected clients
    \FOR{each client \( n \in \mathcal{S}_k \) \textbf{in parallel}} \label{line:client_loop}
        \STATE Initialize \( \omega_{n,k}^0 \gets \omega_k \)
        \FOR{local step \( t = 0 \) to \( T-1 \)} \label{line:local_loop}
            \STATE Compute gradient with noise: \( g_{n,k}^t = \nabla f_n(\omega_{n,k}^t) + \xi_{n,k}^t \) \label{line:gradient}
            \STATE Compute noise-aware weight: \( x_{n,k}^t = \exp(-\gamma \, \xi_{n,k}^t) \) \label{line:weight}
            \STATE Update local model using spodaric and personalized learning in \eqref{eqnlabelnew1} \label{line:update}
        \ENDFOR
        \STATE Client sends \( \omega_{n,k}^T \) to server
    \ENDFOR
    \STATE Server aggregates models using model regularization \eqref{eq: fedavg_personalized}
\ENDFOR
\STATE \textbf{Return:} Optimal global model \( \omega_K \)
\end{algorithmic}
\end{algorithm}}

\begin{figure*}[ht!]
    \centering
    \footnotesize
    \begin{subfigure}[t]{0.245\linewidth}
        \centering
        \includegraphics[width=\linewidth]{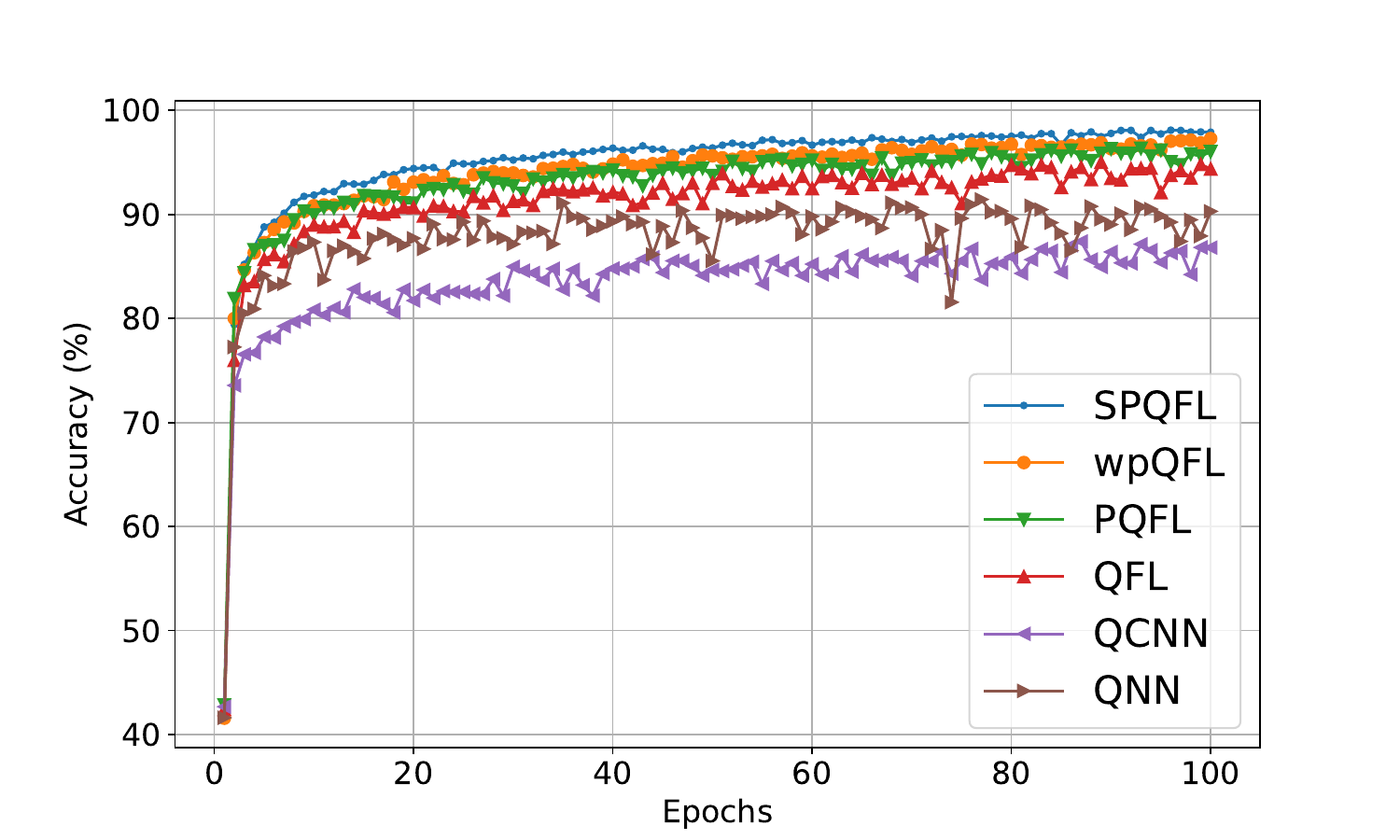}
        \caption{\footnotesize MNIST accuracy.}
    \end{subfigure}
    \hfill
    \begin{subfigure}[t]{0.245\linewidth}
        \centering
        \includegraphics[width=\linewidth]{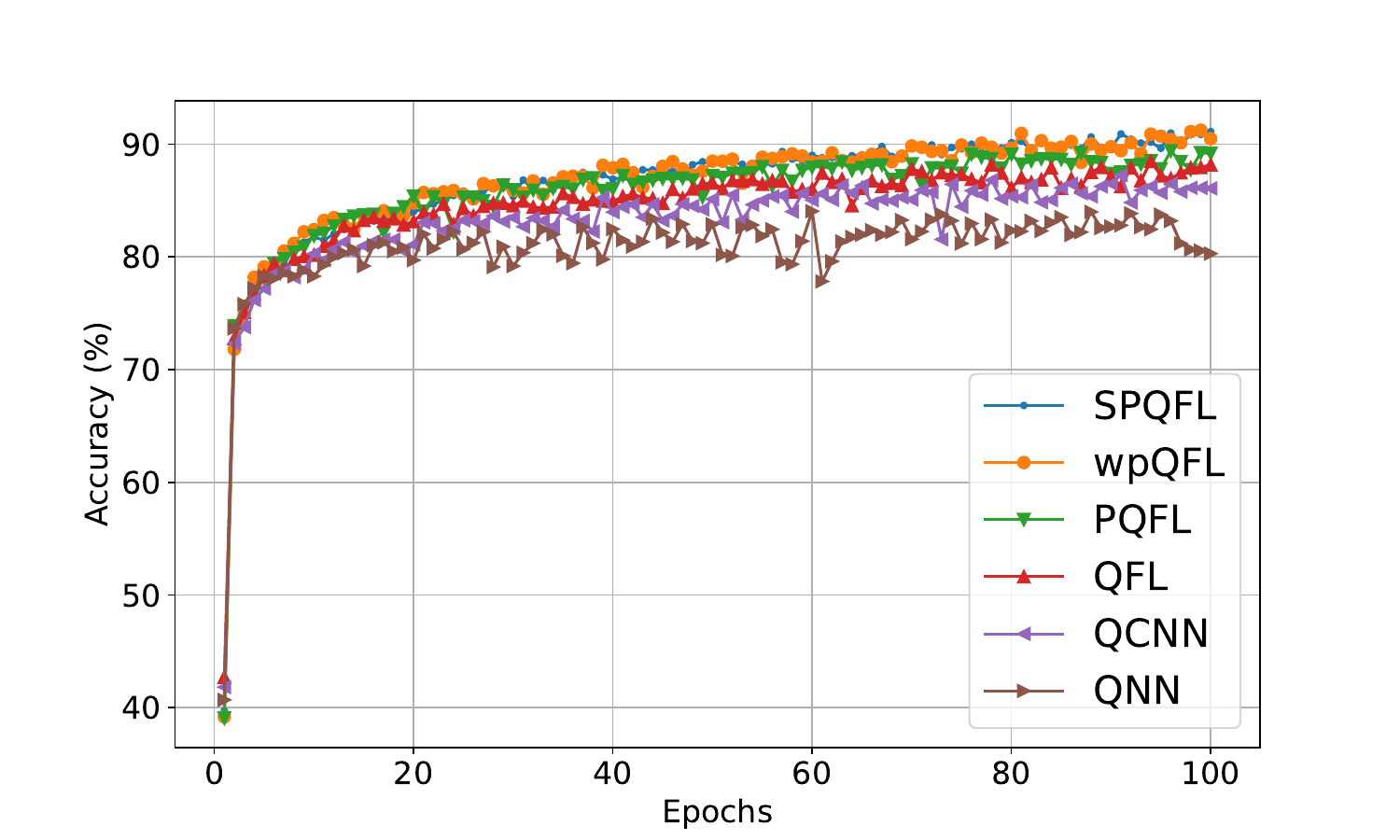}
        \caption{\footnotesize FashionMNIST accuracy.}
    \end{subfigure}
    \hfill
    \begin{subfigure}[t]{0.245\linewidth}
        \centering
        \includegraphics[width=\linewidth]{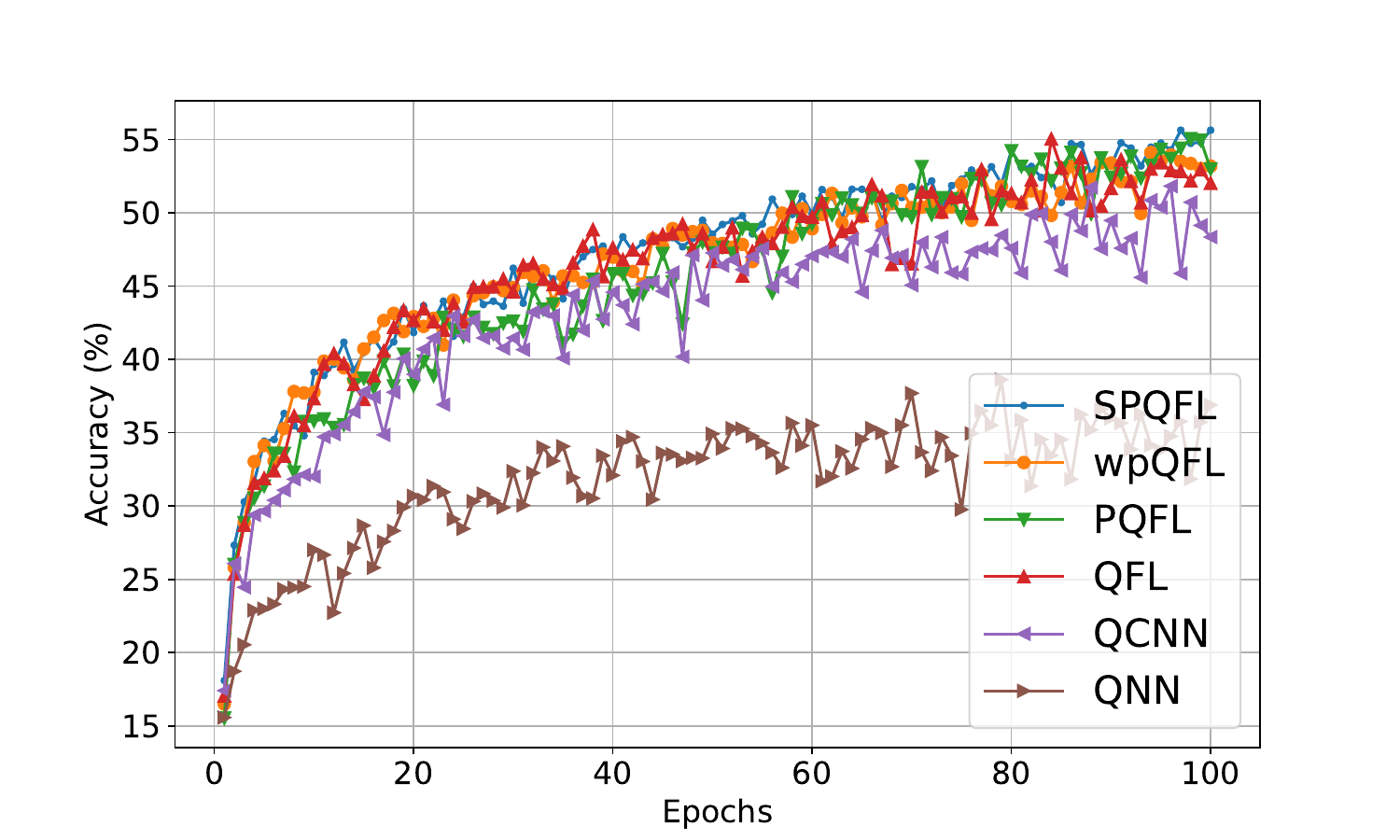}
        \caption{\footnotesize CIFAR-100 accuracy.}
    \end{subfigure}
    \hfill
    \begin{subfigure}[t]{0.245\linewidth}
        \centering
        \includegraphics[width=\linewidth]{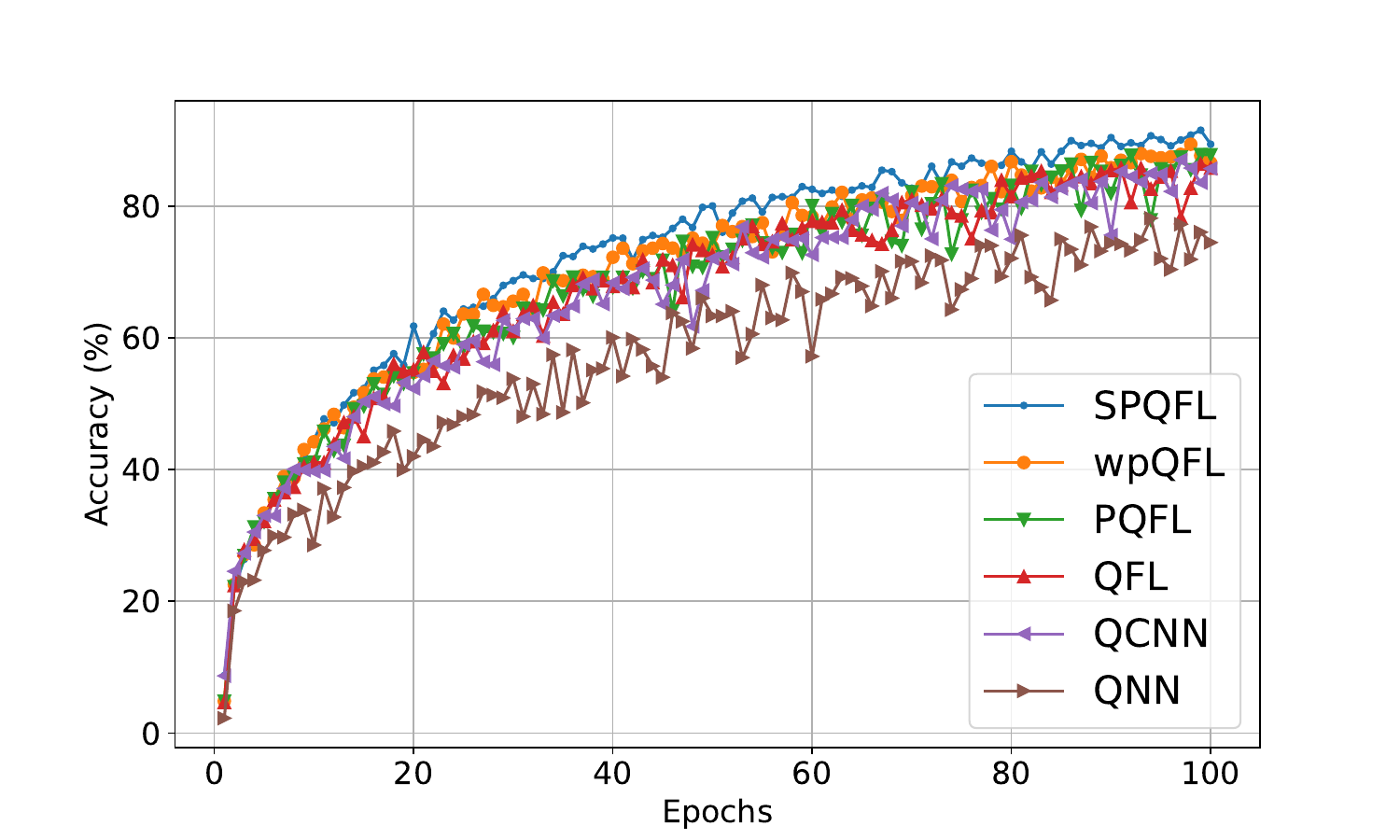}
        \caption{\footnotesize Caltech-101 accuracy.}
    \end{subfigure}

    \vspace{1em} 

    \begin{subfigure}[t]{0.245\linewidth}
        \centering
        \includegraphics[width=\linewidth]{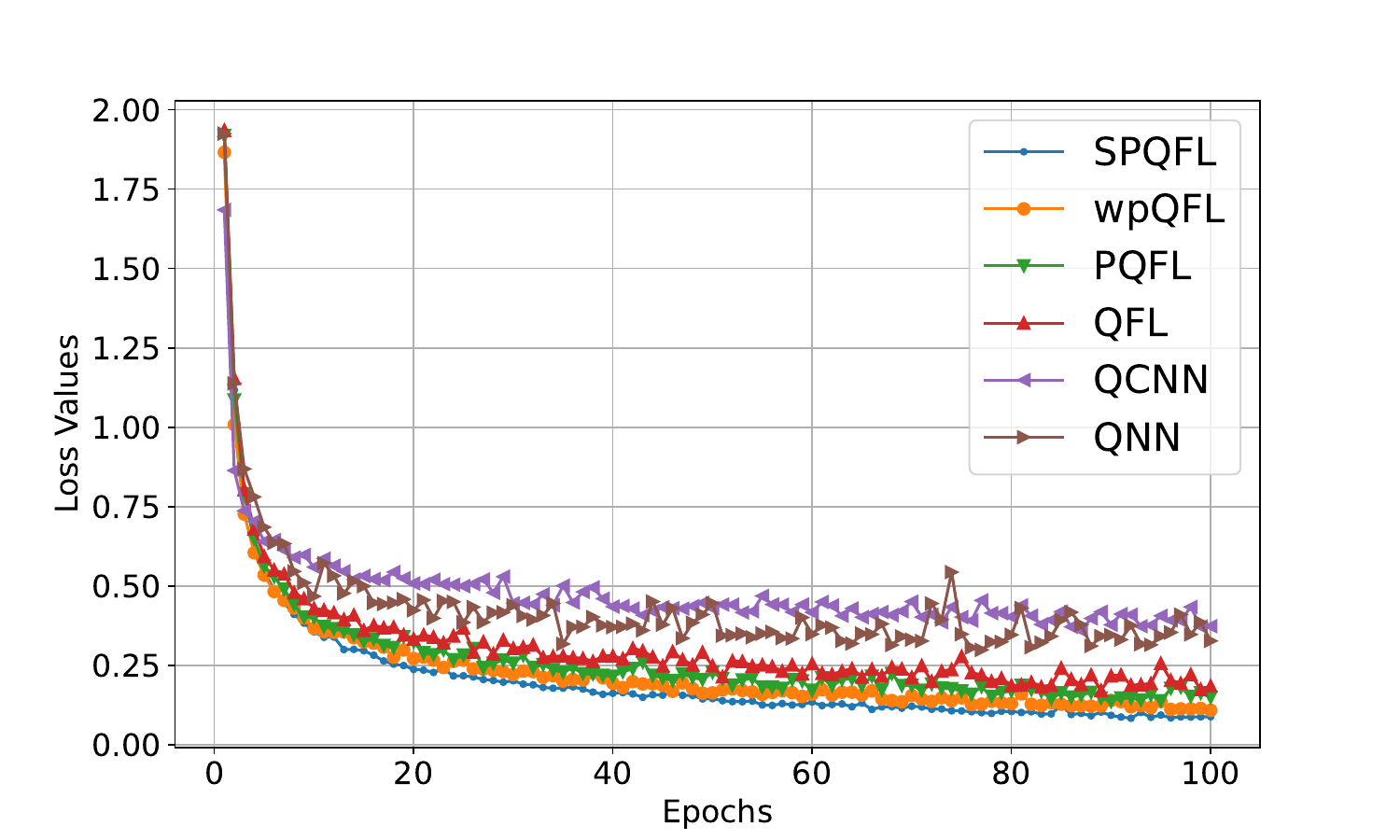}
        \caption{\footnotesize MNIST loss.}
    \end{subfigure}
    \hfill
    \begin{subfigure}[t]{0.245\linewidth}
        \centering
        \includegraphics[width=\linewidth]{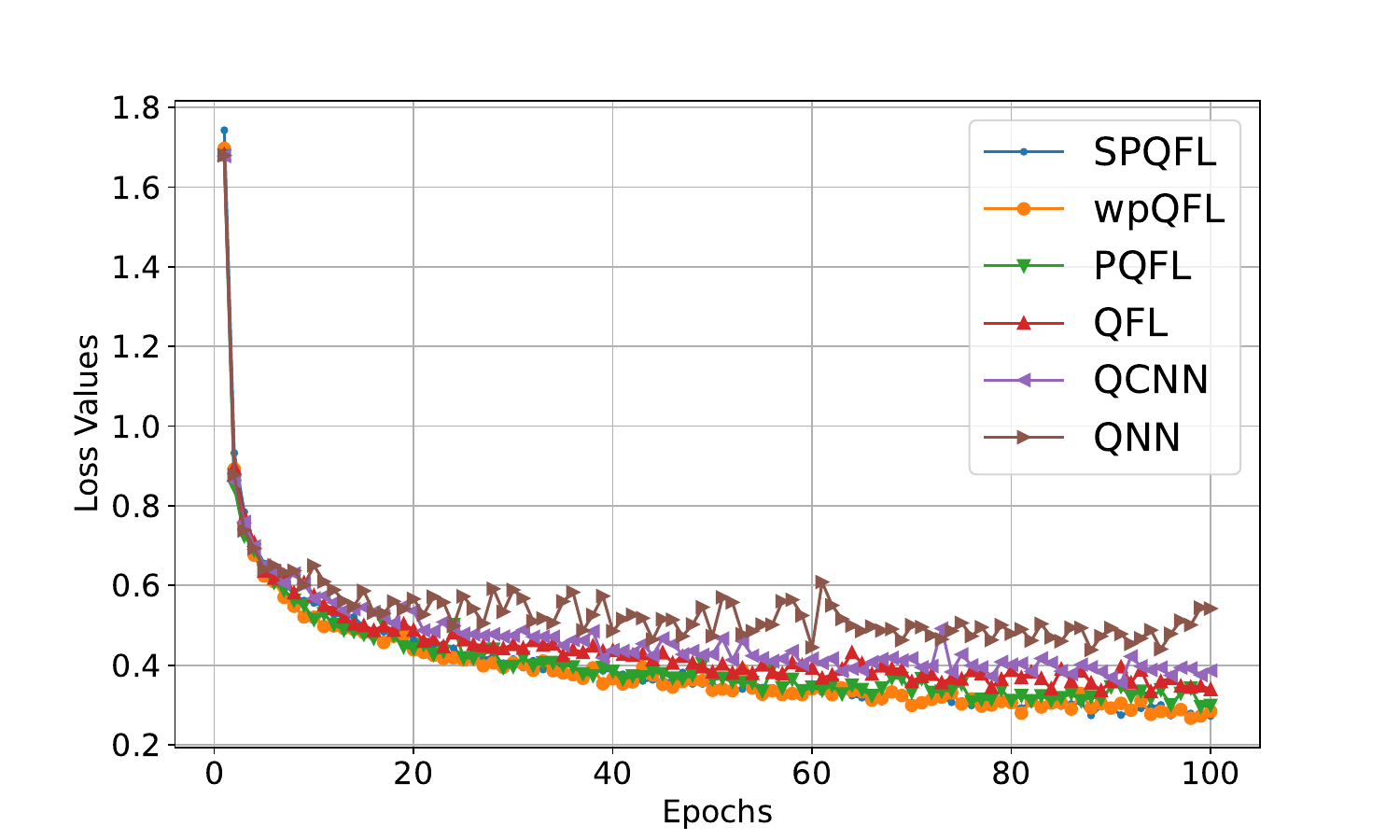}
        \caption{\footnotesize FashionMNIST loss.}
    \end{subfigure}
    \hfill
    \begin{subfigure}[t]{0.245\linewidth}
        \centering
        \includegraphics[width=\linewidth]{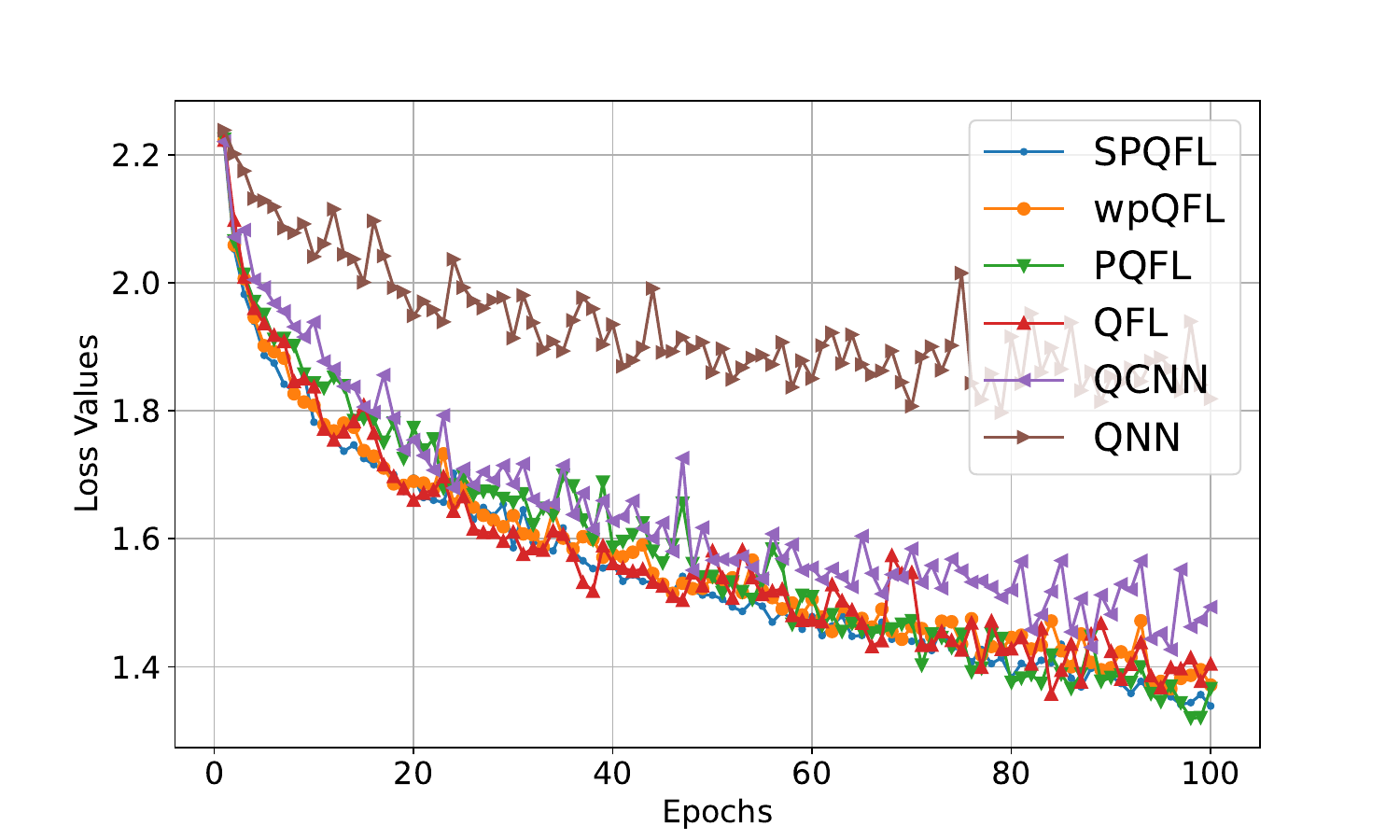}
        \caption{\footnotesize CIFAR-100 loss.}
    \end{subfigure}
    \hfill
    \begin{subfigure}[t]{0.245\linewidth}
        \centering
        \includegraphics[width=\linewidth]{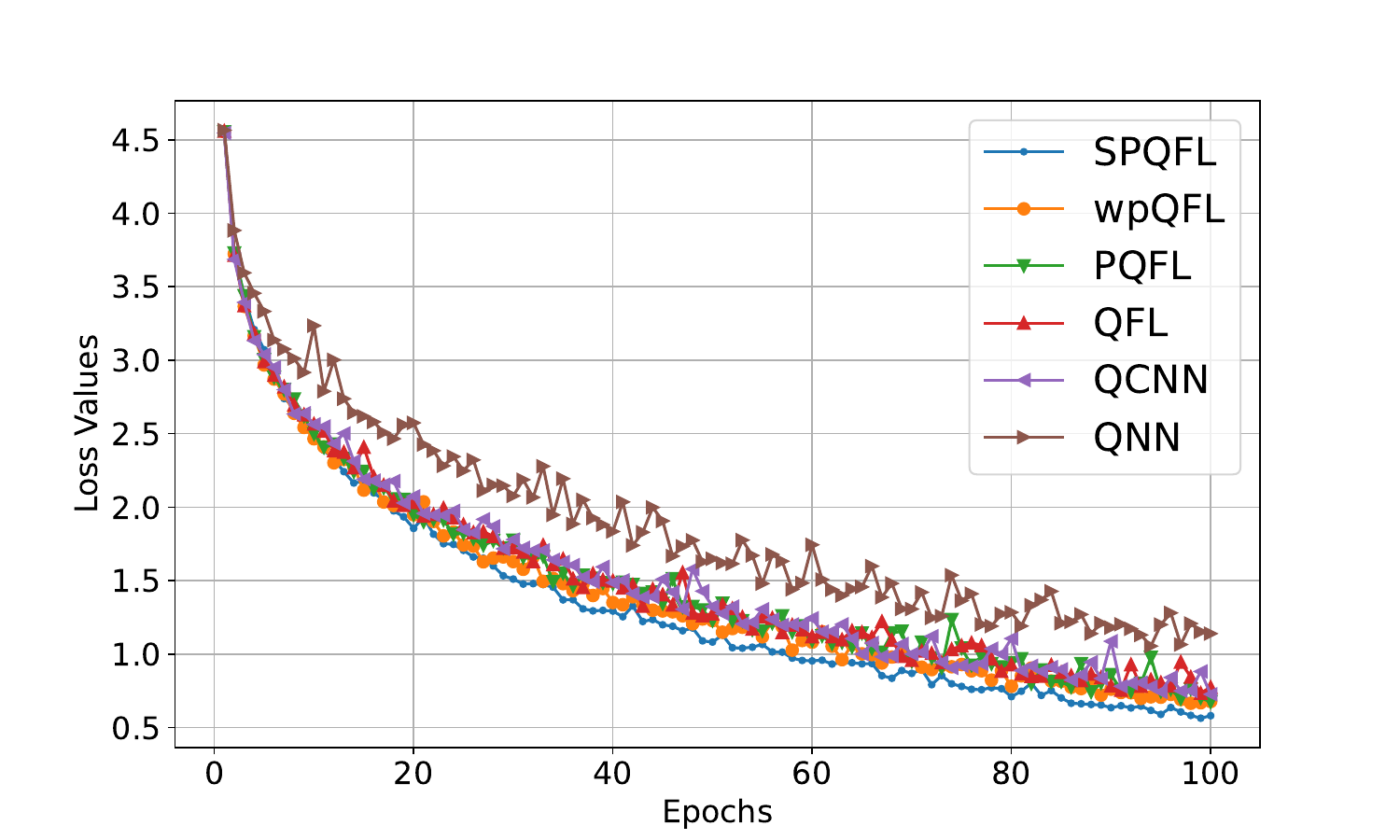}
        \caption{\footnotesize Caltech-101 loss.}
    \end{subfigure}

    \caption{Comparison of SPQFL with other state-of-the-art approaches across datasets. 
    Top row: classification accuracy; bottom row: cross-entropy loss.}
    \label{fig:finalcomp}
\end{figure*}

\subsection{Sporadic Personalized QFL Approach}
Each quantum device in the suggested \textit{SPQFL} framework converts its local data into a quantum model utilizing the complete quantum layer and QNN.  The three distinct components of the QNN model are decoding, quantum circuits, and encoding.  First, qubits and encoding procedures are used to convert classical data to quantum data.  Next, random quantum circuits made up of quantum gates and quantum layers process the quantum data.  Lastly, key characteristics and significant data in the decoding are used to build the local model.  For complicated and large-scale datasets, the encoding and decoding enable the quantum model to provide a local model with appropriate features and to offer an easier-to-process model. To improve model performance, we additionally incorporate regularization-based tailored QFL and occasional learning.  In particular, we assess the modified local model prior to forwarding it to the server, taking into account quantum noise in local model updates.  The server only receives devices that meet the necessary performance level.  If not, the device model is returned to the device for more local epochs and is not transmitted to the server.  The integrity and efficacy of the global model are maintained by this selective approach, which ensures that sub-par models do not make it to the server.

\subsection{Illustrative Results}
We compare our approach to existing state-of-the-art approaches in quantum environments in Fig.~\ref{fig:finalcomp}. For comparison, we choose QNN~\cite{schuld2014quest} for basic QML structure, QCNN~\cite{oh2020tutorial} for the hybrid QML approach, QFL~\cite{chehimi2022quantum} for basic QFL framework, PQFL~\cite{shi2024personalized} for personalized QFL framework, wpQFL~\cite{gurung2024personalized} for weighting averaging based personalized QFL, and our \textit{SPQFL} algorithm. 
\textcolor{black}{Our proposed \textit{SPQFL} outperforms existing approaches in accuracy and convergence speed across multiple datasets.  The benefits come from two design choices.  Regularization-based that penalizes overfitting in local quantum circuits, stabilizes training for heterogeneous clients, and reduces variation in aggregated updates.  Second, the \emph{sporadic (occasional) learning mechanism} prevents noisy or sub-par updates from polluting the global model. Local models are only sent to the server if their validation accuracy is above a predetermined threshold; otherwise, they undergo extra local refining.  This selective participation mitigates the detrimental impact of devices with high decoherence rates or low data quality, leading to more reliable aggregation.}
\textcolor{black}{The performance is evaluated using two complementary metrics: (i) test accuracy (classification rate on the held-out 20\% test set) and (ii) cross-entropy loss.  To ensure robustness, we present the results averaged over three independent runs.  Our sporadic method improves accuracy by $1.6\%$ compared to PQFL in all datasets.  \textit{SPQFL} improves MNIST by 3.03\%, FashionMNIST by 2.51\%, CIFAR-100 by 3.71\%, and Caltech-101 by 6.25\% compared to the regular QFL baseline. The result is also similar for the loss (cross-entropy loss) function. These results demonstrate the adaptability of \textit{SPQFL} across both small- and large-scale datasets, confirming its usefulness as a scalable and noise-resilient framework for QFL.}

\section{Conclusion and Open Research Topics}
\textcolor{black}{In this paper, we have thoroughly studied the challenges} of heterogeneity in QFL settings, including both data and system-level heterogeneity that affect the performance of the quantum model in distributed networks.  We \textcolor{black}{have} investigated numerous mitigation measures that aim to improve data coherence and system compatibility, increasing the effectiveness of QFL deployments. Despite these efforts, our findings highlight inherent limits in existing techniques, notably in terms of scalability, resistance to quantum noise, and practical integration of distinct quantum technologies.  These problems highlight the difficulty of implementing quantum models in a federated environment, where shifts in quantum device capabilities and data distributions have a significant influence on overall system performance.
Several interesting open research topics for tackling the heterogeneity issue in QFL are highlighted below
\begin{itemize}
\color{black}
    \item \textit{Scalability and robustness enhancements:} Existing heterogeneous QFL mitigation methods frequently face scalability issues and are extremely sensitive to quantum noise, limiting their use in large-scale quantum networks in practical settings.  Future research should look at more adaptable and noise-resilient algorithms that retain learning efficiency between clients with different quantum capabilities while also preserving privacy and communication efficiency.
    \item \textit{Advanced error mitigation techniques:} Advanced error mitigation is still a major concern, as quantum systems are inherently fragile. Future QFL frameworks must include both quantum error correction and innovative error-aware learning algorithms that perform well in distributed, noisy settings, addressing not just hardware-level errors but also aggregate errors produced during federated training.
    \item \textit{Impact of quantum network dynamics:} Unlike conventional networks, quantum networks are vulnerable to unique dynamics such as decoherence-induced latency, entanglement generation failures, and variable link reliability. Investigating how these quantum-specific network features affect the stability and performance of QFL systems may result in the creation of communication-aware protocols and a more robust architecture.
\end{itemize}

\bibliography{main}
\bibliographystyle{IEEEtran}

\begin{IEEEbiographynophoto}{Ratun Rahman} is a Ph.D. candidate in the Department of Electrical and Computer Engineering at The University of Alabama in Huntsville, USA. His work focuses on machine learning, federated learning, and quantum machine learning. He has published papers in several IEEE journals, including IEEE TVT, IEEE IoTJ, IEEE TBSE, and IEEE GRSL, and conferences, including NeurIPS and CVPR workshops, IEEE QCE, and IEEE CCNC. 
\end{IEEEbiographynophoto}
\begin{IEEEbiographynophoto}{ Dinh C. Nguyen} (Member, IEEE) is an assistant professor at the Department of Electrical and Computer Engineering, The University of Alabama in Huntsville, USA. He worked as a postdoctoral research associate at Purdue University, USA from 2022 to 2023. He obtained the Ph.D. degree in computer science from Deakin University, Australia in 2021. His current research interests include federated machine learning, Internet of Things, wireless networking, and security. He has published over 50 papers (including over 25 first-authored papers) on top-tier IEEE/ACM conferences and journals such as IEEE Journal on Selected Areas in Communications, IEEE Communications Surveys and Tutorials, IEEE Transactions on Mobile Computing, and IEEE Wireless Communications Magazine. He is an Associate Editor of the IEEE Internet of Things Journal, IEEE Open Journal of the Communications Society, and a Lead Guest Editor of IEEE Internet of Things Magazine on the special issue of federated learning for Industrial Internet of Things.  He received the Best Editor Award from IEEE Open Journal of Communications Society in 2023.
\end{IEEEbiographynophoto}
\begin{IEEEbiographynophoto} {Christo Kurisummoottil Thomas} (Senior Member, IEEE)
received the B.S. degree in electronics and communication engineering from the National Institute of Technology, Calicut, India, in 2010, the M.S. degree
in telecommunication engineering from the Indian Institute of Science, Bengaluru, India, in 2012, and
the Ph.D. degree from EURECOM, France, in 2020. From 2012 to 2014, he was a Staff Design Engineer on 4G LTE at Broadcom Communications,
Bengaluru. From 2014 to 2017, he was a Design Engineer at Intel Corporation, Bengaluru. From November 2020 to June 2022, he was a Staff Engineer on 5G modems at the Wireless Research and Development Division, Qualcomm Inc., Espoo, Finland. He is currently an assistant professor at the Department of Electrical and Computer Engineering, Worcester Polytechnic Institute, USA. Prior to that, he was a Post-Doctoral Fellow with the Department of
Electrical and Computer Engineering, Virginia Tech, during 2022-2025. His research interests
include semantic communications, statistical signal processing, and machine learning for wireless communications. He was a recipient of the Best Student Paper Award from the IEEE SPAWC 2018, Kalamata, Greece. He also received third prize for his team titled “Learned Chester” ML5G-PHY channel estimation challenge, as part of the ITU AI/ML in 5G challenge, conducted at NCSU, USA, in 2020.
\end{IEEEbiographynophoto}

\begin{IEEEbiographynophoto}
    {Walid Saad} (Fellow, IEEE) received the Ph.D.
degree from the University of Oslo, Norway,
in 2010. He is currently a Professor with the
Department of Electrical and Computer Engineering,
Virginia Tech, where he leads the Network Science,
Wireless, and Security (NEWS) Laboratory. He is
also the Next-G Wireless Faculty Lead with Virginia
Tech’s Innovation Campus. His current research
interests include wireless networks (5G/6G/beyond),
machine learning, game theory, security, UAVs,
semantic communications, cyber-physical systems,
and network science. From 2015 to 2017, he was named the Stephen O. Lane
Junior Faculty Fellow with Virginia Tech. In 2017, he was named the College
of Engineering Faculty Fellow. He was a recipient of the NSF CAREER
Award in 2013, the AFOSR Summer Faculty Fellowship in 2014, and the
Young Investigator Award from the Office of Naval Research (ONR) in 2015.
He received the Dean’s Award for Research Excellence from Virginia Tech
in 2019. He was the (coauthor) of 11 conference best paper awards at IEEE
WiOpt in 2009, ICIMP in 2010, IEEE WCNC in 2012, IEEE PIMRC in
2015, IEEE SmartGridComm in 2015, EuCNC in 2017, IEEE GLOBECOM
in 2018 and 2020, IFIP NTMS in 2019, and IEEE ICC in 2020 and 2022.
He was a recipient of the 2015 and 2022 Fred W. Ellersick Prize from the
IEEE Communications Society, the IEEE Communications Society Marconi
Prize Award in 2023, and the IEEE Communications Society Award for
Advances in Communication in 2023. He was also the coauthor of the articles
that received the IEEE Communications Society Young Author Best Paper
Award in 2019, 2021, and 2023. He also received the 2017 IEEE ComSoc
Best Young Professional in Academia Award, the 2018 IEEE ComSoc Radio
Communications Committee Early Achievement Award, and the 2019 IEEE
ComSoc Communication Theory Technical Committee Early Achievement
Award. He was also an IEEE Distinguished Lecturer from 2019 to 2020.
He currently serves as an Area Editor for
IEEE TRANSACTIONS ON NETWORK SCIENCE AND ENGINEERING and
IEEE TRANSACTIONS ON COMMUNICATIONS. He is the Editor-in-Chief
of IEEE TRANSACTIONS ON MACHINE LEARNING IN COMMUNICATIONS
AND NETWORKING.
\end{IEEEbiographynophoto}

\end{document}